\documentclass[aps,prb,twocolumn]{revtex4-1}
\usepackage{bm}
\usepackage{amssymb,amsmath}
\usepackage{graphicx}
\usepackage{times}
\usepackage{color}

\newcommand{\ket}[1]{|{#1}\rangle}
\newcommand{\ketbra}[1]{|{#1}\rangle\langle{#1}|}
\newcommand{\bra}[1]{\langle{#1}|}
\newcommand{\inner}[2]{\langle{#1}|{#2}\rangle}
\DeclareMathOperator*{\Tr}{Tr}
\def\ih{\frac{i}{\hbar}}
\def\ddt{\frac{d}{dt}}

\DeclareMathOperator{\kelvin}{{\rm K}}
\newcommand{\be}{\begin{equation}}
\newcommand{\ee}{\end{equation}}
\newcommand{\bae}{\begin{eqnarray}} \newcommand{\eae}{\end{eqnarray}}
\newcommand{\sandwich}[3]{\left \langle #1 | #2 | #3 \right \rangle}

\numberwithin{equation}{section}

\allowdisplaybreaks[1]

\begin{document}

\title{An analytical continuation approach for evaluating emission lineshapes of molecular aggregates and the adequacy of multichromophoric F\"orster theory}


\author{Leonardo Banchi$^{1}$}
\author{Gianluca Costagliola$^{1,3}$}
\author{Akihito Ishizaki$^{2}$}
\author{Paolo Giorda$^{1}$}

\affiliation{$^1$ Institute for Scientific Interchange Foundation, Via Alassio 11/c 10126 Torino, Italy}
\affiliation{$^2$Institute for Molecular Science, National Institutes of Natural Sciences, Okazaki 444-8585, Japan}
\affiliation{$^3$Dipartimento di Fisica, Universit\'a di Torino, I-10125 Torino, Italy}

\begin{abstract}
In large photosynthetic chromophore-protein complexes not all chromophores are coupled strongly, and thus the situation is well described by formation of delocalized states in certain domains of strongly coupled chromophores.
In order to describe excitation energy transfer among different domains without performing extensive numerical calculations, one of the most popular techniques is a generalization of F\"orster theory to multichromophoric aggregates (generalized F\"orster theory) proposed by Sumi [J. Phys. Chem. B {\bf 103}, 252 (1999)] and Scholes and Fleming [J. Phys. Chem. B {\bf 104}, 1854 (2000)].
The aim of this paper is twofold.
In the first place, by means of analytic continuation and a time convolutionless quantum master equation approach, a theory of emission lineshape of multichromophoric systems or molecular aggregates is proposed.
In the second place, a comprehensive framework that allows for a clear, compact and effective study of the multichromophoric approach in the full general version proposed by Jang, Newton and Silbey [Phys. Rev. Lett. {\bf 92}, 218301 (2004)] is developed.

We apply the present theory to simple paradigmatic systems and we show on one hand the effectiveness of time-convolutionless techniques in deriving lineshape operators and on the other hand we show how the multichromophoric approach can give significant improvements in the determination of energy transfer rates in particular when the systems under study are not the purely F\"orster regime.
The presented scheme allows for an effective implementation of the multichromophoric F\"orster approach which may be of use for simulating energy transfer dynamics in large photosynthetic aggregates, for which massive computational resources are usually required.
Furthermore, our method allows for a systematic comparison of multichromophoric F\"oster and generalized F\"orster theories and for a clear understanding of their respective limits of validity.
\end{abstract}

\date{\today}
\maketitle

\section{Introduction}

The refinement of laser spectroscopic techniques in past few decades has dramatically improved the possibility of studying those systems that are at the basis of the presence of Life on Earth.
\cite{Fleming:1997wz}
Indeed, the main photosynthetic chromophore-protein complexes have been in past decades structurally characterized at the molecular and atomic level and their ability of realizing almost perfect energy transfer processes is currently fostering huge research efforts aimed at unveiling the fundamental mechanisms at the basis of their remarkably high quantum efficiency.
\cite{vanAmerongen:2000tb,Blankenship:2002th}
More recently, the presence of exquisitely quantum mechanical effects in the energy transfer process has been revealed in different photosynthetic complexes even at physiological temperatures.
\cite{Brixner:2005bi,Engel:2007hb,Lee:2007hq,Ishizaki:2009kya,Calhoun:2009bn,Womick:2009km,Collini:2010fy,Panitchayangkoon:2010wk,Lewis:2012vq,SchlauCohen:2012dn,Wong:2012jd,Westenhoff:2012fi}
The observations of long-lived electronic quantum coherence in photosynthetic excitation energy transfer (EET) raised questions about the role of the protein environment in protecting this coherence and the significance of the quantum coherence in light harvesting efficiency.
In order to elucidate the origin of the long-lived quantum coherence and its functional roles in EET processes, much theoretical efforts have been devoted to construct theories to describe quantum dynamics of photosynthetic EET.
\cite{OlayaCastro:2008wd,Mohseni:2008wt,Plenio:2008wn,Jang:2008vf,Rebentrost:2009tj,Caruso:2009vg,Ishizaki:2009uh,Cao:2009cc,Strumpfer:2009jc,Nemeth:2010vm,Mancal:2010ts,Sarovar:2010hs,Tao:2010gq,Huo:2010wx,Hoyer:2010vs,Chin:2010uy,Prior:2010tc,Wu:2010tw,Scholes:2011iq,Turner:2011ef,giorda2011,Christensson:2011ht,Renaud:2011do,Nalbach:2011cr,Ishizaki:2011cx,Kelly:2011dj,Ritschel:2011ht,OlayaCastro:2011hw,Moix:2011vy,Scholak:2011ti,Kreisbeck:2011vm,Olbrich:2011dh,Turner:2012hk,Fassioli:2012gd,Brumer:2012ib,Ishizaki:2012kfa,Berkelbach:2012dl,Hoyer:2012fg,Shim:2012wx,Shabani:2012ir,Jang:2012fi,Chang:2012ef,Sarovar:2013iu}
Generally, quantum dynamic simulations of large and complex molecular systems require massive computer resources, and it is sometimes difficult to obtain physical insights into them even if such extensive simulations are doable.
In large photosynthetic chromophore-protein complexes, however, not all chromophores are coupled strongly but the situation is well described by formation of delocalized states in certain domains of strongly coupled chromophores.
The standard Redfield theory \cite{Redfield:1957tc} or the modified Redfield theory \cite{Zhang:1998ve,Yang:2002vm,Renger:2003wm} are employed to describe energy relaxation within the domains; however, a theory is needed for describing excitation energy transfer between the exciton states in different domains.
For this purpose, F\"orster theory \cite{Forster:1948vy} has been extended to describe this transfer, namely the generalized F\"orster theory (GFT).
\cite{Sumi:1999ws,Mukai:1999hg,Scholes:2000hu,Scholes:2001da}
The original theory must be modified in order to include the details of the complexes; in particular, when the distance between donor and acceptor aggregates is comparable with their physical size one has to properly account for the interaction between excitonic states belonging to different subunits.
Subsequently, the theory has been extended \cite{Jang:2004tt}
in order to account the intra-subunits quantum coherence; within this framework the distinction in different subunits becomes possible when the intra electronic couplings among the chromophores belonging to a subunit $J_{\rm intra}$ is much greater than the electronic couplings $J_{\rm inter}$ between chromophores belonging to different subunits, and when the strength of the system-bath interaction measured by the reorganization energy  $\lambda$ is such that $J_{\rm inter}\ll \lambda \lesssim J_{\rm intra}$.
This approach is termed the multichromophoric F\"orster theory (MCFT).
It may represent a simplified but effective theory for the description of energy transfer processes taking place in large photosynthetic natural and artificial
\cite{RogachJMaterChem2009, CrookerPRL2002,RogachNanoToday2011,emelianova20102d,collini2009coherent,GierschnerPCCP2012,megow2011mixed,buhbut2010built} complexes where the computational efforts required by the full quantum simulations may be prohibitive.
Despite these potential advantages MCFT has been applied in its full generality only in very limited cases and in general its original simplified version, the generalized F\"orster theories (GFT), is usually preferred.
\cite{Raszewski:2008uc}
Furthermore, a general discussion on the limits of validity of MCFT and GFT is still missing, and this is mainly due to the difficulty of expressing the theory within a compact framework.

The main goal of the present paper is to provide such a general and effective framework.
The present analysis is based on analytic continuation and second order perturbative time convolutionless (TCL2) quantum master equation techniques for the evaluation of lineshape operators, \cite{Shibata:1977cw,Renger:2002wv} which are the main tools of the theory.
The application of these methods allows for a variety of fundamental and interesting results that are organized as follows:
In Sec.~\ref{s.rates} we give a brief review of the MCFT and of the main problems that one has to face in order to evaluate the inter-subunits transfer rates.
In Section \ref{s.lineshape} we first review the formalism for the derivation of the exact master equations for the emission lineshape operators based on analytic continuation and we derive (Sec.~\ref{s.detbalance}) a general detailed balance condition for the multichromophoric case.
In Sec.~\ref{s.tcl} we derive the second order perturbation time convolutionless (TCL2) general formalism of MCFT and we show how its limits of validity can be discussed on the basis of physical arguments.
In Sec.~\ref{s.genralizedForstertheory} we briefly discuss generalized F\"orster theory and we show how one can derive such simplified description starting from our general framework.
In Sec.~\ref{s.adequacy} we apply our theory to simple but relevant multichromophoric systems: on one hand we discuss the adequacy of TCL2 techniques in deriving lineshape operators, and on the other hand we study the limits of validity of the full TCL2 MCFT and compare it with an accurate example of GF theory.
Section \ref{s.conclusion} is devoted to concluding remarks.

\section{Multichromophoric transition rates: preliminaries}
\label{s.rates}

Multichromophoric F\"orster theory is a second order perturbation theory with respect to the electronic coupling between different aggregates that can be developed starting with the Frenkel exciton Hamiltonian of the overall photosynthetic complex.
In particular, we consider the following form of electronic excitation and bath Hamiltonians:
\begin{align}
  	H_{eb} &= \sum_N H_{eb}^N + \sum_{NM} H^{NM},
  	\\
  	H_{eb}^N &= H^N + H_b^N + V_{eb}^N,
	\label{e.HebN}
\end{align}
where
\begin{align}
  	H^N
	&=
	\sum_n \epsilon^N_n \ketbra{N,n} + \sum_{nm} J^N_{nm} \ket{N,n}\bra{N,m},
	\\
  	H^{NM}
	&=
	\sum_{nm} J^{NM}_{nm} \ket{N,n}\bra{M,m}
\end{align}
model the electronic Hamiltonian of subunit $N$ and the interaction Hamiltonian between subunits $N$ and $M$, respectively.
We use capital indices for labeling aggregates and lowercase letters for labeling chromophores inside the aggregate:  $\ket{N,n}$ is the excited state of chromophore $n$ inside aggregate $N$, and $\epsilon^N_n$ is its site energy (the Franck-Condon energy).
In the energy eigenstate (exciton) basis the Hamiltonian for the aggregate $N$ reads
\begin{align}
  	H^{N} = \sum_k \hbar \omega^N_k\ketbra{\varepsilon^N_k}
\end{align}
and $U^N_{nk} = \inner{N,n}{\varepsilon^N_k}$ is the change of basis matrix.
The states $\ket{\varepsilon^N_k}$ represent the excitons of subunit $N$ which are delocalized amongst
the aggregate's chromophores $\ket{N,n}$.
Within each subunit the coupling to the bath degrees of freedom is modeled as
\begin{subequations}
\begin{align}
  	V_{eb}^N
	&=
	\sum_n \ketbra{N,n} \, u^N_n \equiv \sum_n V^N_n \, u^N_n,
  	\\
  	H_b^N
	&=
	\sum_n H_b^{N,n}
\end{align}
\label{e.sysbath}
\end{subequations}
where $\sum_N H_b^{N} = \sum_\xi \hbar \omega^b(\xi)\left[p(\xi)^2+q(\xi)^2\right]/2$ model a set of harmonic oscillators,  $p(\xi)$ and $q(\xi)$ are dimensionless coordinates, and $\omega^b(\xi)$ is the frequency of the phonon mode $\xi$.
The bath operators $u^N_n$ are defined by $u^N_n = -\sum_\xi\hbar\omega^b(\xi) d(\xi)^N_n q_\xi$ where $d(\xi)^N_n$ is the dimensionless displacement of the equilibrium configuration of the $\xi$th phonon mode between the ground state and the excited electronic state of the $n$th site in the $N$th aggregate.
The reorganization energy associated with the $n$th pigment is characterized with the displacements as $\lambda^N_n = \sum_\xi \hbar \omega^b(\xi)[d(\xi)^N_n]^2/2$.
It is assumed that $[H_b^{N}, H_{eb}^M] = [V_{eb}^N, H_{eb}^M] = 0$ for $N\neq M$ i.e., the baths of different subunits are independent.

The MC approach consists of an electronic excitation hopping model among aggregates. As shown in Appendix \ref{a.mastereq}, MCFT can be derived from the full quantum evolution of the exciton-phonon system by:
i) applying second order perturbation techniques with respect to $H^{NM}$, based on the assumption that $J^{NM}_{nm} \ll J^N_{nm}, \lambda$,
and
ii) assuming that the donor aggregate $D$ equilibrates on time scales which are much shorter than those characterizing the transfer to the acceptor $A$.
The first approximation permits to obtain a classical-like master equation for the  aggregates populations, with multichromophoric rates.
The equilibration assumption  allows to simplify the problem in that the rates are independent on the actual donor's initial excited state; and the non-equilibrium dynamical features of the donor subunits are thereby neglected.
The resultant multichromophoric (MC) rates read
\begin{align}
	k^{A \leftarrow D}
  	&=
  	\frac{2}{\hbar^2}\Re\int_0^\infty dt \,
	\Tr\left[ A^A(t) J^{AD}  F^D(t)^\dagger J^{DA} \right]
  	\label{e.mcrate_t}
\\
  	&=
  	\frac{1}{2\pi\hbar^2}\int_{-\infty}^\infty d\omega \,
	\Tr\left[ \hat A^A(\omega) J^{AD} \hat F_D(\omega) J^{DA} \right]
  	\label{e.mcrate_w}
\\
 	&\equiv
  	\sum_{aa'dd'}
	\frac{J^{AD}_{ad}J^{AD}_{a'd'}}{2\pi\hbar^2}
  	\int_{-\infty}^\infty d\omega \,
	\hat A_{a'a}^A(\omega) \hat F^D_{dd'}(\omega).
\end{align}
In the above expressions we used the following notation for Fourier transformed operators
$\hat f(\omega) = \int_{-\infty}^{\infty} dt\, e^{i\omega t} f(t)$ and we have defined the MC lineshape operators (LSO) as
\begin{align}
 	A^N_{aa'}(t) &= \bra{N,a}
  	\Tr_b \left [e^{-\ih H^N_{eb} t}\,\openone\otimes
  	\rho^N_b\,e^{\ih H^N_b t}\right]\ket{N,a'},
  	\label{e.A}
  	\\
  	F^N_{dd'}(t) &= \bra{N,d} \Tr_b \left [e^{-\ih H^N_{eb} t}\, \rho^N_{eb}
  	\,e^{\ih H^N_b t}\right]\ket{N,d'},
  	\label{e.F}
\end{align}
where the properties $A^N(t)^\dagger = A^N(-t)$ and $F^N(t)^\dagger = F^N(-t)$ are satisfied.
$\Tr_b \hat{O}$ indicates the partial trace of an operator $\hat{O}$ over the bath degrees of freedom.
In the compact expressions, Eqs.~\eqref{e.mcrate_t} and \eqref{e.mcrate_w} a matrix multiplication between the lineshape operators and the matrix $J^{AD}$ with elements $J^{AD}_{ad}$ is understood.
The two thermal states are defined by $\rho_{eb}^N \propto e^{-\beta H_{eb}^N}$ and $\rho_b \propto e^{-\beta H_{b}^N}$, where $\beta$ is the inverse temperature and the normalization is understood.

The above expressions are the starting point of the MC approach.
In order to assess its validity there are two orders of problem to confront.
One set of problems is related to the theoretical techniques employed for the evaluation of the LSOs.
Explicit expressions are known for monomers \cite{Mukamel:1995us,May:2004vy} but, in general cases, Eqs.~\eqref{e.A} and \eqref{e.F} can not be exactly evaluated because of the interaction, mediated by the bath, between the aggregate's excitons.
In order to proceed further one has to  employ perturbative techniques with respect to the system bath interaction, simulating the dynamics of the operators $A^N(t)$  and $F^N(t)$ via non-Markovian quantum master equations.
One of the standard techniques is a second-order perturbative time-convolution (TC2) quantum master equation.
However, this approach has fatal flaws, even for monomers, in producing absorption lineshapes in
the non-Markovian regimes corresponding to natural situations.
\cite{Ishizaki:2009uh,Ishizaki:2010fx}
It is thus obviously outside the bounds of possibility to apply the TC2 approach to multichromophoric situations.
One is therefore forced to look for a different approach, and this naturally leads to the application of  the second-order perturbative time-convolutionless (TCL2) method, which is able to reproduce accurate absorptive lineshape of a monomer.

The other subtle point that has to be taken into account is the very difference between absorption and emission LSOs. Owing to the equilibration hypothesis of the donor subunit, the emission LSO in Eq.~\eqref{e.F} has an ``initial state'' $\rho^N_{eb}\neq \openone\otimes\rho^N_b$.
This introduces further complications as the electronic degrees of freedom are correlated
with the bath.
In the TCL2 approach to the LSOs developed in the present paper this issue is treated in a simple way exploiting an identity which relates absorption and emission lineshape: $F(t)$ is obtained from $A(t)$ owing to the analytical continuation identity $F(t) \propto A(t-i\hbar\beta)$. This point will be extensively discussed in Sec.~\ref{s.lineshape}.

The second set of problems that one has to face in order to asses the limits of validity and the accuracy of the MC approach is the comparison with the alternative, simpler techniques that are used in the vast majority of the literature. In general, the basis used for the expression of $k^{A \leftarrow D} $ is irrelevant, however one can express the multichromophoric LSOs in Eqs.~\eqref{e.A} and \eqref{e.F} in the exciton basis and see that the distinctive feature of MCFT lies in the presence of off-diagonal terms:
\be
  	A^N(t) = \sum_k A^N_{k,k}(t) \ket{\varepsilon^N_k}\bra{\varepsilon^N_k} +  \sum_{k\neq k'} A^N_{k,k'}(t)\ket{\varepsilon^N_k}\bra{\varepsilon^N_{k'}}
  	\label{e.lsdiag}
\ee
where $A^N_{k,k'}(t)=\sandwich{\varepsilon_k^N}{A^N(t)}{\varepsilon_{k'}^N}$ and and analogously for $F^N(t)$. Correspondingly one can write $k^{A \leftarrow D} = k_{\rm diag}^{A \leftarrow D} + k_\textrm{off-diag}^{A \leftarrow D}$ and the off-diagonal contribution is meant to take into account the complex intra-subunit dynamics.
In the literature typically only the diagonal terms are considered, and this choice leads to the generalized F\"orster rates
\begin{align}
	k_{\rm diag}^{A \leftarrow D}
	&=
	\sum_{k\ell}
	\frac{( \tilde J^{A_k,D_\ell} )^2}{2\pi\hbar^2}
	\int_{-\infty}^\infty d\omega \,
	\hat A^A_{k,k}(\omega) \hat F^D_{\ell,\ell}(\omega),
	\label{e.ratesumi}
\end{align}
where $\tilde J^{A_k,D_\ell}= \sum_{ad} U^A_{ak}\,J^{AD}_{ad}\,U^D_{d\ell}$, and $U^A$, $U^D$ are the acceptor and donor unitary operators that allow to diagonalize the respective electronic Hamiltonians.
The usual F\"orster theory is included as a particular case:
in Ref. \cite{Sumi:1999ws} it has been shown that the multichromophoric rate $k^{A\leftarrow D}$ reduces to the  F\"orster rate $k^{A_k\leftarrow D_\ell}$ when only two specific optically allowed exciton states $A_k$ and $D_\ell$ have non-vanishing average electrostatic interactions $\tilde J^{A_k,D_\ell}$ while all the other $\tilde J^{A_h,D_m}, \, (h,m)\neq(k,\ell)$ are vanishingly small, e.g. in the reaction center of purple bacteria.
Another example of diagonal theory has been employed for describing the rapid excitation energy transfer between two circular aggregates, B800 and B850, in Light Harvesting Complex 2 (LH2) of purple bacteria.\cite{Sumi:1999ws,Mukai:1999hg,Scholes:2000hu,Scholes:2001da}

While the above approximation is widely used, the framework developed in our paper will allow us to investigate the role of the off-diagonal terms in determining the overall transfer rate.
In particular, our benchmark technique will be a diagonal approach based on TCL equations that is in many aspects more accurate than the ones used in the literature (Sec.\ref{s.genralizedForstertheory}).

\section{Evaluation of the lineshape operator}
\label{s.lineshape}

In this Section we lay down the general theory for the evaluation of the LSOs.
We derive the exact expressions for the lineshapes operators and this will be the starting point for the subsequent derivation and study  of the second order approximations in the TCL picture.

Let us consider the absorption lineshape and, in order to simplify the notation, let us remove the index of the aggregate.
A master equation for the lineshape operator has been derived in Ref.~\cite{Jang:2004tt} using projection operator perturbative techniques.
Here we derive formally exact expressions exploiting the Gaussian nature of the bath fluctuations.
Setting $\chi^A(t) = e^{-i H_{eb} t/\hbar} \,\openone\otimes\rho_b\,e^{i H_b t/\hbar}$ then in the ``asymmetric interaction picture'' defined by
$\chi^A_{\rm I}(t) = e^{i (H+H_b)t/\hbar} \,\chi^A(t)\,e^{-i H_b t/\hbar}$
the following equation of motion is satisfied
\begin{align}
	\dot \chi^A_{\rm I}(t) = -\ih V_{eb}(t)\,\chi^A_{\rm I}(t)
	\label{e.noliuv}
\end{align}
yielding
\begin{align}
  	\chi^A_{\rm I}(t)
	=
	\mathcal T_+ \,
	e^{-\ih \int_0^t ds\, V_{eb}(s)} \,\chi^A(0),
\end{align}
where $V_{eb}(t) = e^{i (H+H_b)t/\hbar} \,V_{eb}\,e^{-i (H+H_b)t/\hbar}$ and $\mathcal T_+$ is the standard time ordered product, i.e. the Dyson series.
It is important to stress that Eq.~\eqref{e.noliuv} is not a Liouville equations, as the operator $V_{eb}(t)$ acts only to the right.
The absorption LSO follows from Eq.~\eqref{e.noliuv} by tracing out the bath degrees of freedom.
When the system-bath coupling in Eq.~\eqref{e.sysbath} is considered, the Gaussian property of the phonon bath yields the operators $u_n$ to satisfy the Wick theorem.
Hence, proceeding along the same lines of Ref.~\cite{Ishizaki:2009uh}  we find
\begin{align}
  	A_{\rm I}(t)
	&=
	\mathcal T_+
	e^{-\frac1{\hbar^2} \sum_{aa'} \int_0^t
  	ds\int_0^sds'\, C_{aa'}(s-s') V_a(s)\, V_{a'}(s')} A(0),
  	\label{e.AgausI}
\end{align}
where $ C_{aa'}(t-s) = \Tr\left[\rho_b\,u_a(t)\,u_{a'}(s)\right]$ is the bath correlation function.
Despite not being essential, in this paper, we assume that the bath acts independently on the different chromophores as usual, i.e.
\begin{align}	
	C_{aa'}(t)
	=
	C(t) \,\delta_{aa'}.
	\label{e.indipbath}
\end{align}
The bath correlation function can be expressed in terms of the spectral density
$J(\omega)$ as
\begin{align}
	C(t)
	=
	\frac\hbar\pi
	\int_{-\infty}^{\infty} d\omega
	\frac{J(\omega)}{e^{\beta\hbar\omega}-1} e^{i\omega t},
  	\label{e.CJ}
\end{align}
where $J(\omega)$ has been extended to negative frequencies, $J(-\omega) = -J(\omega)$.
The spectral density $J(\omega)$ and the reorganization energy $\lambda$ are related with each other via $\lambda = \int_0^\infty d\omega\,J(\omega)/(\pi\omega)$.
The absorption LSO is then
\begin{align}
	A(t)
	=
	e^{-\ih Ht}\mathcal T_+ e^{-\frac1{\hbar^2}
	\sum_{a} \int_0^tds\int_0^sds'\, C(s-s') V_a(s)\, V_{a}(s')}.
	\label{e.Agaus}
\end{align}
From Eq.~\eqref{e.AgausI} we obtain an equation of motion for the LSO that, in the interaction picture, reads
\begin{align}
  	\ddt  A_{\rm I}(t) = \int_0^tds\,\mathcal T_+
	\left[ K(t, s) \, A_{\rm I}(t)\right],
	\label{e.formaleqmot}
\end{align}
where the dissipation kernel is
\begin{align}
   K(t, s)
   =
   -\frac1{\hbar^2}\,
   C(t-s)\sum_a V_a(t) \,V_a(s)~.
  \label{e.kern}
\end{align}
It is important to stress that Eq.~\eqref{e.formaleqmot} is formally exact provided that the expansions in Eqs.~\eqref{e.Agaus} and \eqref{e.kern} are inserted in the time ordered integral.  The latter however mixes the operators entering in Eqs.~\eqref{e.Agaus} and \eqref{e.kern} making Eq.~\eqref{e.formaleqmot} not suitable for practical calculations.

Trying to proceed similarly for deriving the emission lineshape operator leads to complications as  the Wick theorem cannot be applied straightforwardly due to excitation-environment correlations corresponding to the Stokes shift
in $\chi^F(0)=F(0)$.
A possible solution to this problem would be to consider the equilibrated state $\rho_{eb}$ as a state resulting from an initially factorized state $\rho_f(t_i) = \rho_e(t_i)\otimes\rho_b(t_i)$  corresponding to the Franck-Condon transition and letting $t_i \rightarrow-\infty$.
Such an approach can be analyzed with the help of time-evolutions of quantum master equations.
However, here we consider a much simplified description.
Our approach consists in exploiting the identity
\begin{align}
   F(t) =  A(t-i\hbar \beta) / Z
   \quad
   \text{with}
   \quad
   Z  = \Tr A(-i\hbar\beta),
  \label{e.idFA}
\end{align}
which can be derived straightforwardly from Eqs.~\eqref{e.A} and \eqref{e.F}.
The analytic continuation identity permits to tackle the unfactorized initial condition in a simple way.
In the frequency domain the above identity reads
\begin{align}
  \hat F(\omega)
   =  \frac{e^{-\beta\hbar\omega}}{Z}\,\hat A(\omega)
   \quad
   \text{with}
   \quad
  Z = \int d\omega\,\frac{e^{-\beta\hbar\omega}}{2\pi}\,
  \Tr \hat A(\omega)
  \label{e.idFAw}
\end{align}
and was used also by Sumi in Ref. \cite{Sumi:1999ws} for expressing the emission lineshape in terms of the absorption lineshape.

When the aggregate $N$ consists of a single monomer the above equations, Eqs.~\eqref{e.Agaus} and \eqref{e.idFA} can be evaluated exactly without any approximations:
\begin{align}
   A_{\rm mon}(t)
   &= e^{-i \epsilon t/\hbar}e^{- g(t) },
\\
   F_{\rm mon}(t)
   &= e^{-i (\epsilon-2\lambda) t/\hbar} e^{- g(t)^*},
  \label{e.monomer}
\end{align}
with
\begin{align}
	g(t) = \frac1{\hbar^2} \int_0^tds\,\int_0^sds' \, C(s')
\end{align}
being the line-broadening function. \cite{Mukamel:1995us}
The emission lineshape directly follows from the absorption lineshape and Eq.~\eqref{e.idFA}.
Indeed, it is straightforward to show
\begin{align}
	g(t-i\hbar\beta) = g(t)^*-\ih 2\lambda t - \beta \lambda.
	\label{e.idg}
\end{align}

In the subsequent section we derive approximated expression for the lineshapes by
making use of lowest meaningful order expansion, the second order, of the TCL approach.
The latter allow to account for all specific cases of system-bath interactions and their performances will be compared in the relevant situations in Sec.~\ref{s.adequacy}.

\subsection{Multichromophoric detailed balance condition}
\label{s.detbalance}

Before analyzing the TCL derivation of LSO we discuss how, owing to the analytical continuation identity described above, the detailed balance condition (DBC) can be generalized to the multichromophoric scenarios. In the standard DBC the ratio $k^{D \leftarrow A} / k^{A\leftarrow D}$ only depends on the temperature and on the energy difference between the donor and the acceptor.
On the other hand, it is possible to show that when the donor or the acceptor are composed of aggregates, DBC depends also on the details of aggregate, as the intra-aggregate electronic interactions, and on the parameters of the bath.
The multichromophoric detailed balance condition naturally follows by inserting the analytic continuation identity in the frequency domain, Eq.~\eqref{e.idFAw}, into the definition of the MC rate in Eq.~\eqref{e.mcrate_w}. It is straightforward to show that
\begin{align}
  \frac{k^{A\leftarrow D}}{k^{D\leftarrow A}} = \frac{Z^A}{Z^D}
  \equiv \frac{\Tr[e^{-\beta H^A_{eb}]}}{\Tr[e^{-\beta H^D_{eb}}]}
  \label{e.detbalance}
\end{align}
where $H^N_{eb}$, for $N=A,D$, is the Hamiltonian in Eq.~\eqref{e.HebN} of the aggregate $N$ which models both the electronic interaction amongst chromophores and the coupling
with the environment.
The latter identity in Eq.~\eqref{e.detbalance} follows by using the relation between the partition functions $Z^N$ defined in Eq.~\eqref{e.idFA} and \eqref{e.idFAw}, together with the definition of $A^{N}(-i\hbar\beta)$.
As for the general features, the above equation does not depend on the electronic interaction amongst the aggregates: it is given by the ratio of the partition function of the two aggregates so that it relies only on the details of the equilibrated state of the single subunits. The latter however, as discussed in Section \ref{s.tcl}, can be different from the Gibbs state, especially for higher values of $\lambda$, and depends also in a non-trivial way on the parameters of the bath.

\subsection{Time-convolutionless formalism }\label{s.tcl}

In this section we present the main result of this paper.
As described below the application of TCL techniques allows for a neat formal description of the multichromophoric approach that is on one hand compact and general and on the other hand it allows to pinpoint the main physically relevant issues at the basis of the approach.
Both the equilibrated state of the donor subunit  and the absorption and emission lineshapes can be determined with equations that are formally consistent. Furthermore, one can separate the off-diagonal contribution from the diagonal one, and its relevance can be studied in terms of the electronic structure of a given subunit and of the specific bath properties (the details of the derivation can be found in Appendix (\ref{a.tcl2}) where the general theory of TCL approach for determining the LSOs is sketched - together with an analogous derivation with TC2 method).

Within a second order time-convolutionless (TCL2) framework the absorption
LSO can be written as (see Appendix \ref{a.tcl2})
\begin{align}
	A^{\rm TCL}(t)
	=
	a_{ex}(t) \cdot a(t).
	\label{e.Atclfact}
\end{align}
In Eq.~\eqref{e.Atclfact}, $a_{ex}(t)$ is defined as
\begin{align}
	a_{ex}(t)&=e^{-i Ht/\hbar}\,e^{-g(t)},
\end{align}
whereas
$a(t)$ is the matrix solution of the following differential equation:
\begin{align}
  	\dot a(t) = -O(t) a(t) ,
	\label{e.adiffeq}
\end{align}
where the matrix $O$ is given by
\begin{align}
  	\begin{aligned}
	O_{kk'}(t)
	&=
	\frac{1}{\hbar^2}
    e^{i\omega_{k,k'}t}
  	\int_0^t ds\,  C(s)\,\Xi_{kk'}(s),
\\
  	\Xi_{kk'}(t)
	&=
	\sum_{k''} \Xi_{k,k'}^{k''}
  	\left(e^{-i \omega_{k'',k'} t}-1\right),
  	\end{aligned}
  	\label{e.Lope}
\end{align}
with $\omega_{k'',k'} = \omega_{k''}-\omega_{k'}$ and $\Xi_{k,k'}^{k''}=\sum_a U_{ak}U_{ak''}^2U_{ak'}$.
Similarly, owing to the analytic continuation identity in Eqs.~\eqref{e.idFA} and \eqref{e.idg}
we find
\begin{align}
  	F^{\rm TCL}(t)
	=
	f_{ex}(t) \cdot f(t),
  	\label{e.Ftclfact}
\end{align}
where $f_{ex}(t)$ is given as
\begin{align}
 	f_{ex}(t)
	=
	e^{-\beta H}\,e^{-i (H-2\lambda)t/\hbar}\,e^{-g^*(t)}
\end{align}
and
$f(t)$ is the matrix solution of the differential equation,
\begin{align}
	\dot f(t) = -O(t-i\hbar\beta)f(t)
\end{align}
with the initial condition
\begin{align}
	f(0) = a(-i\hbar\beta)/\Tr A^{\rm TCL}(-i\hbar\beta).
	\label{e.Feqstate}
\end{align}
The latter is obtained by numerical integration of Eq.~\eqref{e.adiffeq} in the imaginary time.

Equation~\eqref{e.Atclfact} states that $A^{\rm TCL}(t)$ is given by
the product of a matrix $a_{ex}(t)$ diagonal in the exciton basis and the matrix $a(t)$ that contains
the off-diagonal terms.
This decomposition allows to separate the different contributes of multichromphoric lineshapes and to discuss their physical interpretation.
The $a_{ex}(t)$ factor describes the lineshapes corresponding to excitons of the aggregate:
the $k$th diagonal element is the lineshape corresponding to $k$th exciton with energy $\hbar \omega_k$.
Note that each excitonic lineshape can be determined exactly without any approximation:
the interaction with the bath is taken into account by means of the broadening function $g(t)$.
On the other hand,
the interaction between the excitons, mediated by the bath degrees of freedom, is taken into account by the matrix valued quantity $a(t)$ whose physical interpretation is more intricate.
The matrix operator $O(t)$ depends, on one hand, on the relation between the exciton and site basis through the tensor $\Xi_{k,k'}^{k''}$;
when the resulting matrix $\Xi(t)$ is diagonal so it is $a(t)$, and this in particular happens when site and exciton basis coincide.
On the other hand $O(t)$  depends on the relations between typical frequencies of the system, $\omega_{k'',k'}$, and typical frequencies of the bath that determine the behavior of $C(s)$; in particular if $\omega_{k,k'}$ are smaller than all the typical frequencies of the bath (high-temperature, Markovian case), then $e^{i\omega_{k,k'}t}\simeq 1$: the operator $O(t)$ is essentially null on the relevant time scales of the bath and hence $a(t)$ is close to the identity; in this case the behavior of the aggregate is well approximated by the lineshapes of independent excitons.
When typical frequencies of the aggregate are comparable or greater than the bath's ones, then the details of the aggregate have to be considered and they can give rise to a modification of the diagonal elements or to the occurrence of some off-diagonal terms. Relevant examples are discussed in the next sections.

As for the emission lineshape, we finally note that since the determination of the equilibrated state $F(0)=\rho_{e}^{\rm TCL}=\Tr_b \rho_{eb}$ involves the same formal structure previously described one has:
\begin{align}
	\rho_{e}^{\rm TCL}
	=
	a_{ex}(-i\hbar\beta) \cdot a(-i\hbar\beta)/\Tr A^{\rm TCL}(-i\hbar\beta),
\end{align}
where $a_{ex}(-i\hbar\beta)= e^{-\beta(H-\lambda)}$ and $a(-i\hbar\beta)$ is the solution of $\dot a(-i\hbar\tau) = -O(-i\hbar\tau)\,a(-i\hbar\tau)$ where the integration interval is $[0,\beta=1/k_B T]$. Therefore, the same above discussion holds.
Indeed, the structure of the operator  $O(- i \hbar \tau )$ is the same as before: the details of the aggregate Hamiltonian are encoded in the same tensor $\Xi_{k,k'}^{k''}$, while the ``dynamical'' aspects are determined by the comparison of the relevant frequencies of the system $\omega_{k'' k'}$ with the thermal correlation frequency $\omega_T=1/{\beta\hbar}$.
In particular, when $\omega_{k'' k'} \ll \omega_T$ e.g., high temperature regime, one has that $e^{-\tau \omega_{k k'}} \approx 1$ and the integration interval is  small; therefore $a(-i\hbar\beta)\approx \openone $, the off-diagonal term in $\rho_{e}^{\rm TCL}$ are vanishingly small and the only effect is a renormalization of the exciton energies that in general, and depending on $\lambda$, can lead to an equilibrated state different from the Gibbs state.
In the opposite case, $\omega_{k ''k'} \gg \omega_T$ i.e., in particular at low temperature, the integration interval grows and the off-diagonal terms in $\rho_{e}^{\rm TCL}$ come into play.

The above discussion clearly highlights that within the TCL2 picture developed in this paper, each steps of the MC approach can be easily analyzed by studying a single matrix valued operator $O$ that embeds all the structural and physical features of the given system under observation. Within this picture all further possible and relevant approximations can be discussed in a clear way.

\subsection{Generalized F\"orster Theory}\label{s.genralizedForstertheory}

Most of the literature about electronic energy transfer between molecular aggregates employ generalized F\"orster theory (GFT) for estimating the transition rates between multichromophoric donors and acceptors.
\cite{Sumi:1999ws,Mukai:1999hg,Scholes:2000hu,Scholes:2001da,Raszewski:2008uc,Renger:2011ui}
Sometimes the internal exciton dynamics within the aggregates is included phenomenologically  with the use of Redfield/modified Redfield theories. \cite{Renger:2011ui,Raszewski:2008uc}
However, similar to the assumptions of MCFT, usually it is assumed that the donor has reached its equilibrium state $\rho^e$ before the transfer takes place.
In the latter case, the rate between aggregates is given by the weighted sum of the F\"orster rate $k^{A_k\leftarrow D_j}$ from every exciton $D_j$ of the donor to every exciton $A_k$ of the acceptor, and the weights are given by the the probability of finding the donor aggregate in the exciton $D_j$.
The latter are given by the equilibrated state, which in the simplest description is modeled by a Gibbs state, although this is not accurate because of system-bath interaction, which causes the Stokes shift.\cite{Fleming:1996td,Ishizaki:2009uh}
Within the formalism developed in this paper the equilibrated state is absorbed into the definition of the emission lineshape; in this way one can show that GFT rates can be obtained as a limiting case of the  MCFT rates, provided that the lineshapes are diagonal in the exciton basis, Eq.~\eqref{e.ratesumi}.

Equations \eqref{e.Atclfact}-\eqref{e.Feqstate} are the starting points for implementing various approximations and  thus for deriving different GF expressions in a controlled way. A first way for obtaining a GF expression is to neglect $a(t)$, $f(t)$ and $a(-i \beta \hbar)$ terms in the evaluation of the absorption, emission line shapes. This approximation is a very simple one: the excitons lineshapes can be analytically evaluated, and the equilibrated state is the aggregate Gibbs state.

A second GF expression can be derived following the standard approach for evaluating the lineshapes in GFT, which is based on the off-diagonal elements of a master equation which describes the dynamics within each aggregate. \cite{Renger:2002wv}
In this case TCL2 approximation is usually employed, although with some exceptions, \cite{renger2007primary}
as it reduces to the exact expression for monomers. One can show that the GFT lineshapes can also be obtained by implementing a secular approximation in the general expression, Eq.~\eqref{e.Lope}: oscillating parts in Eq.~\eqref{e.Atcl} are selectively removed and the diagonal and off-diagonal elements are completely decoupled. In particular, in the exciton basis
\begin{align}
	\ddt A^{\rm GF}_{k,k}(t)
	&=
	\left(
		-i \omega_k
		-\frac{1}{\hbar^2}
		\int_0^t ds
		\sum_{k''} \Xi_{k,k}^{k''}\, C(s) \,e^{-i \omega_{k'',k} s}
	\right)
	A^{\rm GF}_{k,k}(t).
	\label{e.Agftcl}
\end{align}
Similar expressions hold for $A^{\rm GF}_{k,k'}(t)$, however, since $A^{\rm GF}_{k,k'}(0) = 0$, one has that $A^{\rm GF}_{k,k'}(t) \equiv 0$ for $k\neq k'$.
In other words, the secular approximation decouples the diagonal terms from the off-diagonal ones, forcing the LSO to be diagonal in the exciton basis.
Equation~\eqref{e.Agftcl} was obtained by Renger and Marcus \cite{Renger:2002wv} with different methods; furthermore, in order to simplify the numerics, they treat in a Markovian way the elements with $k\neq k''$ in \eqref{e.Agftcl}.
Our goal will be to compare the accuracy of MCFT and GFT in describing the dynamics within the aggregates; to this aim we consider the more general expression \eqref{e.Agftcl} without any further approximation and in the following we will refer to it as GF-TCL2.
As for the emission lineshape we implement the analytic continuation identity, Eq.~\eqref{e.idFA} in Eq.~\eqref{e.Agftcl}.
In this way, no other approximations are necessary in the description as the analytic continuation identity automatically considers the effect of the stokes shift and of the equilibrated initial state. We note however that the equilibrated initial states differs from the one obtained with the multichromophoric approach, and this will be more evident at low temperatures.

\section{Adequacy of MultiChromophoric  approach}\label{s.adequacy}

\subsection{Time-convolutionless lineshapes}\label{s.tctcl}

In order to test the adequacy of the MC-TCL2 approach we start by considering simple systems. Our first goal is to compare the  MC-TCL2 rates with those obtainable with other relevant theories widely used in the literature.
\cite{Jang:2008vf,Ishizaki:2009uh,Tao:2010gq,Huo:2010wx,Prior:2010tc,Renaud:2011do,Nalbach:2011cr,Ritschel:2011ht,Kreisbeck:2011vm,Berkelbach:2012dl,Chang:2012ef}
One of such approaches is the second-order cumulant time-nonlocal (2CTNL) quantum dynamics, which takes into account environmental reorganization processes or dynamic Stokes shifts.
\cite{Ishizaki:2009uh}
It implements a hierarchical representation  \cite{Takagahara:1977wq,Tanimura:1989dd,Tanimura:2006ga,Xu:2007ws} for numerical calculations.
The second theory is the time-convolution approach for the evaluation of LSOs.
Although even in the case of single monomer's absorption lineshapes TC2 master equations are not able to reproduce the real lineshapes and give unphysical results,\cite{Ishizaki:2009uh,Ishizaki:2010fx} the only multi-chromophoric analysis available in the literature
\cite{Jang:2004tt,Jang:2003wh,jang:9312} use TC2 approximation for calculating the LSO.

We start by analyzing a simple but relevant model i.e., a trimer whose Hamiltonian is the following:
\begin{align}
  H = \begin{pmatrix}
    200 & 100 & 2
    \\
    100 & 100 & 2
    \\
    2 & 2 & 0
  \end{pmatrix}
  \label{e.Htrimer}
\end{align}
where the values are given in ${\rm cm}^{-1}$.
Since  $J_{\rm intra} = 100\,{\rm cm}^{-1} \gg J_{\rm inter} = 2\,{\rm cm}^{-1}$ the multichromophoric aggregate is clearly formed by a dimer subunit and by a monomer weakly coupled with the dimer. The exciton energies in the dimer are
\begin{align}
  \hbar\omega^{\rm dimer}_+ &\simeq 261.8\,{\rm cm}^{-1}, &
  \hbar\omega^{\rm dimer}_- &\simeq 38.2\,{\rm cm}^{-1}.
  \label{e.dim200100energies}
\end{align}
The values used for this example are typical values that can be found for example in biological systems.
The system-bath interaction is implemented by choosing a  Drude-Lorentz spectral density \cite{Breuer:2002wp}
\begin{align}
  	J(\omega) = 2\lambda \frac{\gamma\omega}{\omega^2+\gamma^2},
  	\label{e.drude}
\end{align}
where $\gamma^{-1}$ corresponds to the timescales of the bath-induced fluctuation-dissipation process.
The reorganization energy $\lambda$ characterizes the strength of its coupling with the excitons.
Moreover, in this section, we consider the system to be at room temperature, $T = 300 \kelvin$.
The electronic coupling between the dimer and the monomer ($J_{\rm inter}$) is chosen to be small in order to justify the second order perturbation theory which is one of the assumption in MCFT.

\begin{figure}[t!]
  \begin{center}
    \includegraphics[width=.5\textwidth]{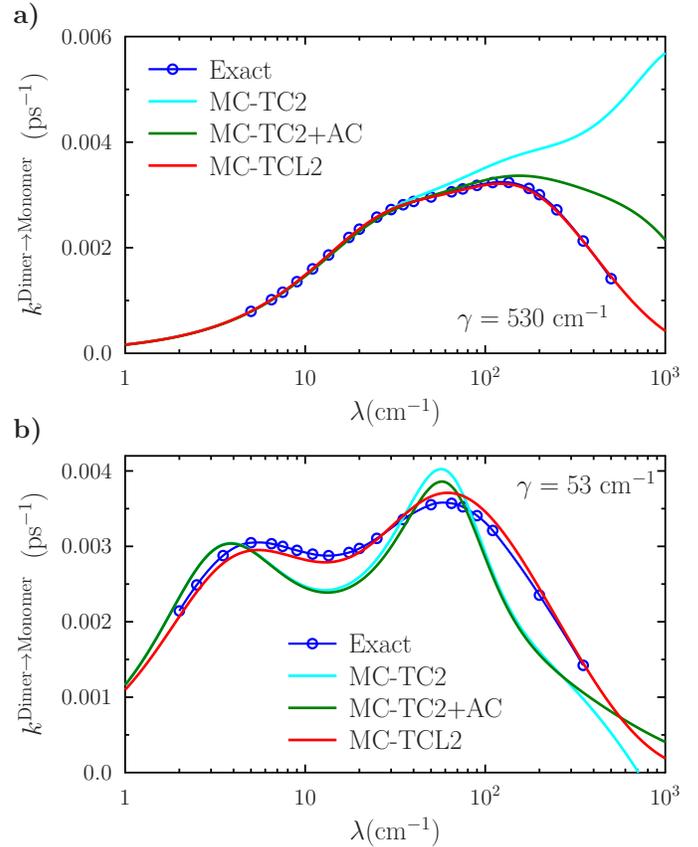}
  \end{center}
  \caption{Multichromophoric rates vs. $\lambda~({\rm cm}^{-1})$
  from the dimer to the monomer for $\gamma = 530\,{\rm cm}^{-1}$ (upper plot)
  and $\gamma = 53\,{\rm cm}^{-1}$ (lower plot). In the above,
  the MC curves are evaluated with the MCFT, where the lineshapes are obtained
  with different theories. In particular, MC TC2 is obtained from Eqs.~\eqref{e.Anz}
  and \eqref{e.Fnz}. In MC TC2+AC, the absorption is still calculated with Eq.~\eqref{e.Anz} while the emission is obtained from Eq.~\eqref{e.Fnzsumi}.
  }
  \label{fig.mescolone2}
\end{figure}

In our discussion we consider the energy transfer occurring from the dimer subunit to the monomer.
Indeed, the multichromophoric donor requires the evaluation of the dimers' emission lineshapes that involve more and relevant approximations.
The back-transfer process from the monomer to the dimer is not considered as it is related with the forward transfer process via the generalized detailed balance condition Eq.~\eqref{e.detbalance}, and thus it does not convey further relevant information.

The multichromophoric rate in Eq.~\eqref{e.mcrate_w} calculated with
time-convolution-less equations are shown in in Fig.~\ref{fig.mescolone2}, for $\gamma = 530\, {\rm cm^{-1}} \gg \omega_\pm$ (Markovian regime) and
$\gamma = 53\, {\rm cm^{-1}}$ (non-Markovian regime).

As anticipated, in addition to the rates given by MCFT we have also performed an exact full quantum calculation by means of 2CTNL approach:
the exact time evolution of the trimer density matrix is obtained and then the rates are estimated with a least-square algorithm.
Furthermore, the rates with time-convolution techniques with or without the use of analytic continuations (see Appendix (\ref{a.tc2})) are shown for comparison.

In the Markovian regime  (upper panel Fig.~\ref{fig.mescolone2}) our simulations show that the TCL2 treatment gives excellent agreement with the accurate full quantum calculations for all values of reorganization energy $\lambda$.
On the other hand the TC2 approaches are accurate only in the weak coupling regime, while in the strong coupling regime they completely fail to reproduce the exact data.
On the other hand, it is important to emphasize that MC-TCL2 approach, despite being a second order approximation, works excellently even in the strong coupling regime.

In the lower panel of Fig.~\ref{fig.mescolone2} are the reported the rates for the case $\gamma = 53\,{\rm cm}^{-1}$, a value which is comparable with the fluctuations timescales obtained in experiments.
Here, the Markovian approximation is not justified anymore, as $\hbar\gamma$ is of the same order of the dimers energies, Eq.~\eqref{e.dim200100energies}.
As in the Markovian case the agreement between the full quantum  and the MC-TCL2 description is extremely good for all values of $\lambda$.

The behavior of the MC-TC2 rates is more complex. The introduction of an analytic continuation approach in the non-Markovian do not lead to significant differences. While the rates evaluated with this theories coincide and qualitatively reproduce the exact one, a close look to the lineshapes in this regime shows that TC2 master equations are not able to give the correct physical description of the LSO: even for small values of $\lambda$ MC-TC2 lineshapes show two peaks. This phenomenon, which becomes more evident for increasing $\lambda$, as already discussed in Ref.~\cite{Ishizaki:2010fx}, is completely unphysical.
On the other hand, this kind of drawback does not affect the lineshapes derived with TCL2 approach for all values of parameters analyzed.
However, the results of this section show that the MCFT approach succeed in determining the correct transfer rates, and a second order treatment of the lineshapes, as given by TCL2, is sufficient.

\begin{figure}[t!]
  \centering
  \includegraphics[width=.5\textwidth]{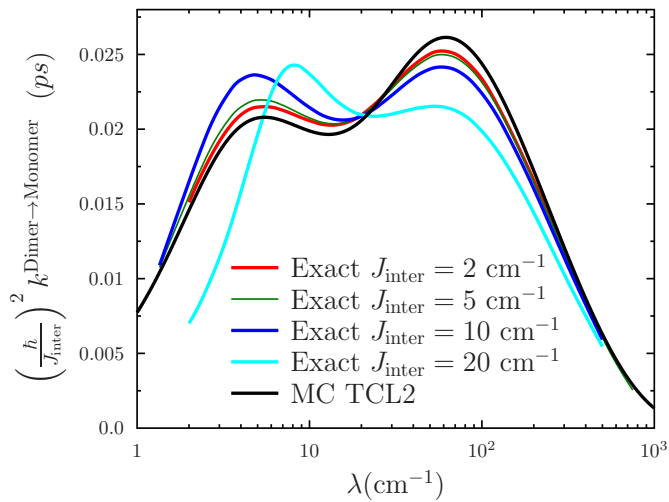}
  \caption{Rescaled rates for the Hamiltonian \eqref{e.dim200100energies} for $T=300\kelvin$  and $\gamma = 53\,{\rm cm}^{-1}$ and different values of $J_{\rm inter}$  }
  \label{fig:ham200100breakdown}
\end{figure}

We conclude this section by analyzing the behavior of the multichromophoric approach when the distinction between subunits is no longer possible.
In particular this is happens when the condition $J_{\rm inter}\ll \lambda$ is violated. In Fig.~\ref{fig:ham200100breakdown} we show the simulations for the normalized rate $\tilde{k}_{D \rightarrow A}= ({\hbar}/{J_{\rm inter}} )^2 k_{D \rightarrow A}$ for the Hamiltonian in Eq.~\eqref{e.dim200100energies} and for growing values of $J_{\rm inter}=J_{13}=J_{23}$.
Owing to the rescaling the MC-TCL2 rates are represented by a single curve and they are compared with the rates obtained with full quantum simulation approach.
While the MC-TCL2 method is always able to capture the rates' qualitative behavior, a quantitative agreement for large $J_{\rm inter}$ is obtained only for high values of $\lambda$.
The result is expected since for growing values of $J_{\rm inter}$ the main hypotheses on which the multichromophoric approach is based are no longer verified, and in particular the dynamics within the dimer do not lead to an equilibration of the subunit on timescales shorter than those characterizing the energy transfer to the acceptor.

\subsection{Multichromophoric vs. generalized F\"orster}
\label{s.mcvsgf}

In order to study the role of the off-diagonal terms and thus describe the differences between MCFT and GFT we consider this simple model of a trimer
\begin{align}
  H = \begin{pmatrix}
    E_1 & J_{12} & J_{13}
    \\
    J_{12} & E_2 & J_{23}
    \\
    J_{13} & J_{23} & 0
  \end{pmatrix}
  \label{e.Htrimer_site}
\end{align}
where the first two sites form a strongly connected dimer, whereas the coupling
between the dimer and the trimer is small i.e., $J_{12} \gg J_{13},J_{23}$. The dimer subunit can be
diagonalized by an orthogonal matrix $U(\theta)$, i.e. a rotation in the energy space
of the pigments in the dimer with some angle $\theta$. In the new basis the
Hamiltonian takes the form
\begin{align}
  \begin{pmatrix}
    \varepsilon+\delta & 0 & \tilde J_+\\
    0 & \varepsilon & \tilde J_- \\
    \tilde J_+ & \tilde J_{-} & 0
  \end{pmatrix}
  \label{e.Htrimer_exc}
\end{align}
As already mentioned the relation between exciton and site basis encoded in $U(\theta)$ determines
the relevance of the off-diagonal terms in the LSOs. Indeed, in the exciton basis the operator in \eqref{e.Lope}
that determines the dynamics of $a(t)$ reads
\begin{align}
	\Xi(t)
  	&=
	\frac14
	\begin{pmatrix}
	2\alpha \sin^2 2\theta	&	\alpha^* \sin 4\theta
	\\
    -\alpha\sin 4\theta	& 	2\alpha^*\sin^2 2\theta
  	\end{pmatrix},
 	 \label{e.xit}
 \\
  	\alpha
  	&=
  	1-e^{-i\delta t}.
\end{align}
In particular the latter is diagonal when $\theta=0$ i.e., the excitons and site basis coincide and each site interacts independently with
its own bath. $\Xi(t)$ is again diagonal  when $\theta=\pi/4$ i.e., when the sites energy are degenerate and
the excitons are given by the symmetric and antisymmetric superposition of site basis states. In this case one can write
the dimer-bath interaction term in the exciton basis in the following way:
\begin{align}
	\sum_{1,2} \ketbra{i}\, u_i
	=
	\openone (b_1+b_2) + \sigma_x (b_1-b2),
\end{align}
where $\sigma_x=\ket 1 \bra 2 + \ket 2 \bra 1$, and consequently one can redefine the baths in terms of the center of mass coordinate $b_+=(b_1+b_2)/\sqrt{2}$ and the difference between the baths coordinates $b_-=(u_1-u_2)/\sqrt{2}$.  Correspondingly one can decompose the dimer's Hamiltonian as $H=H_+ + H_-$, with $H_+= \varepsilon \openone  + \openone \otimes b_+ + H_{b^+}$, $H_-= \delta \ketbra{\varepsilon_+} + \sigma_x\otimes b_- + H_{b^-}$  and $[H_+,H_-]=0$ i.e. the two baths are independent.
Due to symmetry of this specific interaction, the baths are unable to distinguish the excitons. In particular the $\sigma_x$ terms swaps the two excitons and the due to the bosonic properties of the baths the interaction term in Eq.~\eqref{e.Agaus}  ($\sigma_x e^{-i \delta \ketbra{\varepsilon_+} t } \sigma_x$) is proportional to the identity, therefore no off-diagonal terms appears in the LSOs. This fact, due to the bosonic properties of the bath and Wick's theorem, is not specific to second order perturbation but it applies at any order of the perturbation treatment.

Aside from the previous two special cases, in order to describe the relevance of off-diagonal terms one can focus on Eq.~\eqref{e.xit} and fix the conditions for having maximal off diagonal elements.
A necessary (although not sufficient) condition for maximizing the role of the off-diagonal terms is to set $\theta=\pi/8$. See Eq.~\eqref{e.xit}.
The other free parameters are (see Eq.~\eqref{e.Htrimer_exc}: the off-set with respect to the monomer's site energy $\varepsilon$, the difference between the dimer's excitons energies $\delta$, and the electronic couplings $J_-$, $J_+$.
We chose to focus on the following dimer Hamiltonian:
\begin{align}
  \begin{pmatrix}
      253 &  63
      \\
      63 & 126
  \end{pmatrix}
  \label{e.gham1}
\end{align}
where $\delta = 180\,{\rm cm}^{-1}$ and $\varepsilon = 100\,{\rm cm}^{-1}$.
Indeed, as already discussed in section \ref{s.tcl}, the role of the off-diagonal elements becomes important when the typical frequency of the system $\delta/\hbar$ is comparable with the smallest frequency of the bath.
In the non-Markovian high-temperature ($T=300 \kelvin $) regime, the latter corresponds to $\gamma = 53\, {\rm cm}^{-1}/\hbar (10\,{\rm ps}^{-1}) \lesssim \delta$.

\begin{figure}
        \centering
	\includegraphics[width=.45\textwidth]{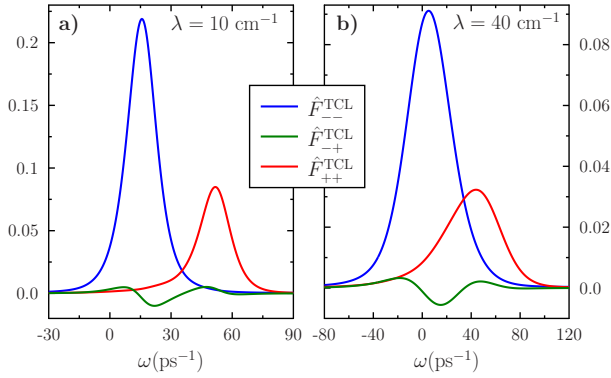}
  \caption{Emission lineshapes for Hamiltonian, Eq.~\eqref{e.gham1}; $T=300K$,
  $\gamma = 53\,{\rm cm}^{-1}$;  $\lambda=10\,{\rm cm}^{-1}$ (a) and $\lambda=40\,{\rm cm}^{-1}$ (b)}
  \label{fig:emission_gham1}
\end{figure}

In terms of emission lineshapes the presence of the off-diagonal terms is revealed in the global MC rate when: $i)$ the monomer's absorption lineshape sufficiently overlaps with the off-diagonal terms; $ii)$ the effect is not screened by the overlap of the monomer's absorption lineshape with lineshape corresponding to the
highest dimer's exciton; this latter point can be achieved if the two excitons' lineshapes are sufficiently separated in frequency, and this is determined by
difference in energy between the excitons $\delta$, and if $\lambda$ is not too high.
Indeed, the emission lineshapes are shown, for $\lambda=10$ and $40\,{\rm cm}^{-1}$ in Fig.~\ref{fig:emission_gham1}, the absorption lineshape (not shown) is centered in zero.
When $\lambda$ grows the lineshapes are more broadened and the monomer's absorption lineshape strongly overlaps with both diagonal terms, thus reducing the relevance of the off-diagonal terms.

\begin{figure}[t!]
	\centering
  	\includegraphics[width=.5\textwidth]{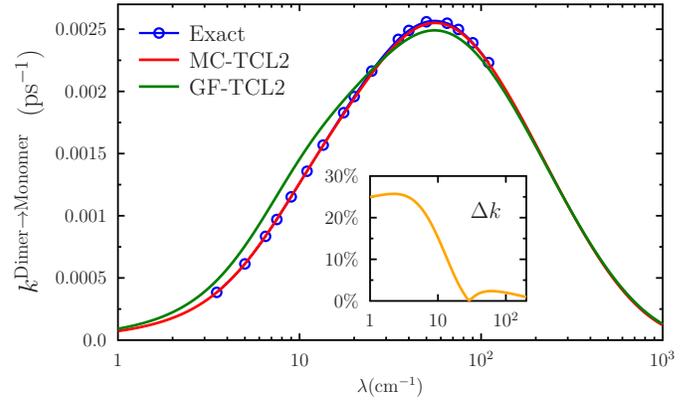}
  	\caption{Rates for the Hamiltonian in Eq.~\eqref{e.gham1} for $T=300\kelvin$ and $\gamma = 53\,{\rm cm}^{-1}$.
  	In MC-TCL the LSO absorption lineshape is given by Eq.~\eqref{e.Atclfact}, in GF-TCL is given by Eq.~\eqref{e.Agftcl}.
	In every case, the emission lineshape is
  	obtained from the respective absorption lineshape owing to the analytic continuation identity,  Eq.~\eqref{e.idFA}.
  	The inset shows the rates' relative errors $\Delta k$ for different values of $\lambda$.
  }
  \label{fig:gham1}
\end{figure}

\begin{figure}[t!]
  	\centering
  	\includegraphics[width=.5\textwidth]{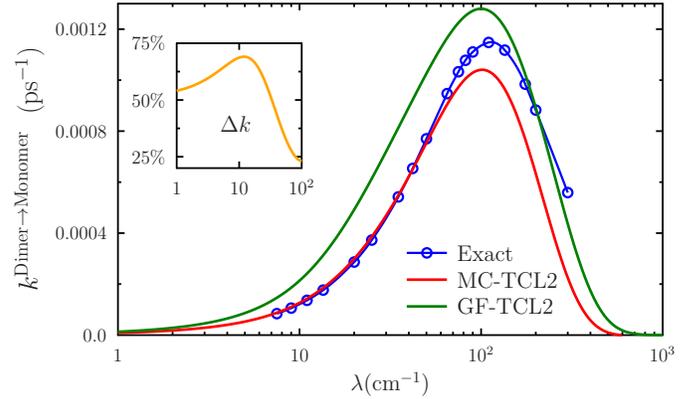}
  	\caption{Rates for the Hamiltonian in Eq.~\eqref{e.gham1} for $T=100\kelvin$ and $\gamma = 53\,{\rm cm}^{-1}$.
  	In MC-TCL the LSO absorption lineshape is given by \eqref{e.Atclfact}, in GF-TCL is given by \eqref{e.Agftcl}.
	In every case, the emission lineshape is obtained from the respective absorption lineshape owing to the analytic continuation identity, Eq.~\eqref{e.idFA}.
  	The inset shows the rates' relative errors $\Delta k$ for different values of $\lambda$.
  }
  	\label{fig:gham3}
\end{figure}

In Fig.~\ref{fig:gham3} we compare the rates obtained with MC-TCL2 approach for the Hamiltonian, Eq.~\eqref{e.gham1} with the ones obtained with full quantum simulations and with the generalized F\"orster theory described in Eq.~\eqref{e.Agftcl} (GF-TCL2) where the lineshapes have been calculated by imposing the decoupling of the diagonal terms.
Our results show that the multichromophoric approach properly reproduces the accurate rates
In general the transfer rates are very small, as we have chosen small electronic couplings amongst the dimer and the monomer ($J_{13}=2\,{\rm cm}^{-1}, J_{23}=0$) in order to maintain the distinction between the two subunits.
The figure of merit able to quantify the relevance of the off-diagonal terms is the rate's relative error $\Delta k = \lvert k_{\rm GF}- k_{\rm MC}) \rvert /k_{\rm MC}$ and it is plot in the inset of Fig.~\ref{fig:gham1}.
We can see that the error between the MC-TCL2 and generalized F\"orster approach is quite relevant up to $\lambda\approx 20\,{\rm cm}^{-1}$; for $\lambda = 10\,{\rm cm}^{-1}$, $\Delta k$ is around $27\%$.
The behavior shown by $\Delta k$ is typical for the Hamiltonians with $\theta\approx\pi/8$ in the high temperature regime: the effects of the off-diagonal terms is in general quite relevant in the ``true'' MC regime i.e.,  when $J_{\rm inter}\ll \lambda \lesssim J_{\rm intra}$ and the second order perturbation theory at the basis of the approach holds.
When $\lambda \gg J_{\rm intra}$ the system enters in a F\"orster regime; as described before, in  this region the presence off-diagonal terms of the LSOs is screened due to the broadening of the excitons lineshapes, therefore  the generalized F\"orster methods are in good agreement with the MC-TCL2.

We finally examine the effect of the temperature. We use the following values $\gamma=53\,{\rm cm}^{-1}$ and $T=100\,{\rm K}$ and the Hamiltonian, Eq.~\eqref{e.gham1}.
For these values of temperature $K_B T \approx 70\,{\rm cm}^{-1}$, it is comparable with $\gamma$ and the typical energies of the system.
The results of our simulations in figure \ref{fig:gham3} show that again we have two distinct regions.
For $\lambda <90\,{\rm cm}^{-1}$ the MC-TCL2 approach is in complete agreement with the accurate rate, while the GF-TCL2 is far less accurate; indeed, as shown in the inset of Fig.~\ref{fig:gham3}, the relative error $\Delta k$ between the two rates is very large up to $\lambda=100$, and it reaches its maximum  $\Delta k=65\%$ at physiological values of the reorganization energy, $\lambda=20\,{\rm cm^{-1}}$.
On the contrary, for very large values of $\lambda$ we see that the best approximation is given by a generalized Forster approach.
This example highlights the limits of validity of the MC-TCL2 approach.
For values of reorganization energy typical of biological systems and when $J_{\rm inter}\ll \lambda \lesssim J_{\rm intra}$ the multichromophoric approach is extremely accurate in determining the transfer rates.
For very high values of $\lambda$ and at low temperature the method at becomes less accurate; in these region of the parameters the equilibrated state of the dimer starts to significantly differ from the actual one;
in particular its off-diagonal terms are orders of magnitude smaller than the correct ones.
Furthermore, and independently from the equilibrated state used, the off-diagonal terms in the emission lineshape equations lead in  to significant errors in the determination of the overall rate. This fact is less relevant as for the back transfer rate (not shown), where the agreement of MC-TCL2 and the accurate result is fairly good.

For the sake of completeness we include the emission lineshapes as given by MC TCL2 for two values of $\lambda$ where the MC TCL2 differs from the accurate rate.

\begin{figure}
       \centering
       \includegraphics[width=.45\textwidth]{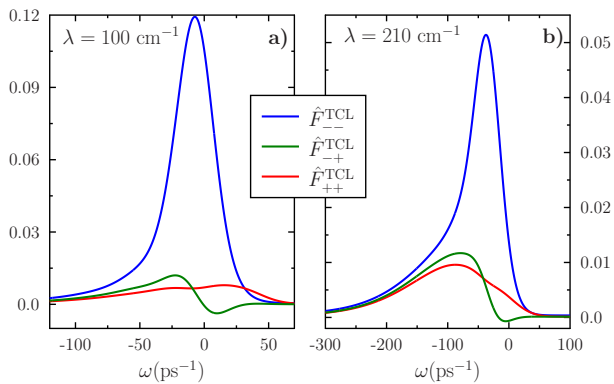}
  \caption{Emission lineshapes for the dimer of  Hamiltonian in Eq.~\eqref{e.gham1} for  $\gamma=53\,{\rm cm}^{-1}$ and $T=100\,{\rm K}$, {$\lambda = 100\,{\rm cm}^{-1}$} (a)
  $\lambda = 210\,{\rm cm}^{-1}$ (b)}.
  \label{fig:lineshapesT100}
\end{figure}

\section{Conclusions}\label{s.conclusion}

The study of energy transfer in large light harvesting complexes can be simplified when these structures can be divided into weakly coupled subunits of strongly interacting chromophores, i.e., when $J_{\rm inter}\ll \lambda \ll J_{\rm intra}$, where $J_{\rm inter (\rm intra)}$ is the inter (intra) subunit electronic coupling and $\lambda$ is the reorganization energy measuring the strength of the system-bath interaction. In this context, the multichromophoric approach \cite{Sumi:1999ws, Jang:2004tt} was put forward to account for the intra-subunits bath mediated coherent interaction in the determination of the incoherent inter-subunit energy transfer rates. In principle, this approach allows to overcome the problem posed by the massive computational resources required by the sophisticated techniques available for the accurate simulation of the quantum dynamics of the whole complexes. However a comprehensive discussion of the approach is still missing.

In this paper we provide a clear, compact and general picture of the multichromophoric approach that allows to study its general scope and limits of validity.
Our discussion is based on the evaluation of lineshape operators with a new approach based on analytic continuation and second order time convolutionless techniques (TCL2). The latter are found to be strictly necessary since the alternative time convolution approach, despite being widely used in the literature, has severe limitations when real biological conditions are considered.

The picture developed allows one to identify a F\"orster like contribution that can be determined without approximations, and that describes the incoherent hopping amongst excitons belonging to different aggregates. On the other hand, the bath mediated coherent interaction between the excitons of a given subunit is taken into account by an independent matrix valued quantity whose off-diagonal terms are peculiar of the multichromophoric theory.

In this way, one can immediately identify, on the basis of the electronic properties of the subunits and of the typical bath timescales, if and which multichromophoric (off-diagonal) terms are relevant and must be taken into account, and to derive simplified versions of the multichromophoric transfer rates that encompass the (diagonal) generalized F\"orster theories used in the literature.

We have applied our general theory to simple prototypical examples of photosynthetic systems composed by a dimer weakly coupled with a monomer. We have considered both Markovian and non-Markovian regimes and we have compared our results with those obtainable with other accurate quantum dynamic simulations and with meaningful instances of generalized F\"orster approximations that can be derived within our approach.
Our analysis shows that the TCL2 multichromophoric theory is able to accurately describe the coherent subunits dynamics and it can outperform the generalized Forster techniques especially when $J_{\rm inter}\ll \lambda \lesssim J_{\rm intra}$. \\
The main limitations of the approach occur in the first place when the subdivision in subunits is no longer possible i.e., $J_{\rm inter} \approx \lambda$ and in the less obvious case of low temperatures and extremely high values of reorganization energy.

In conclusion, our analysis provides a comprehensive and general framework in which the multichromophoric approach to energy transfer processes between complex photosynthetic subunits can be discussed.
While our numerical analysis has focused on simple trimeric structures, the framework developed allows for the study of more complex systems and for the judicious identification of the aggregates and regimes for which a full multichromophoric picture is necessary.

\begin{acknowledgements}
PG and LB would like to thank Seth Lloyd for useful discussions.
AI is grateful for JSPS Grants-in-Aid for Scientific Research (KAKENHI) Grant Number 24850018.
\end{acknowledgements}

\appendix

\section{Multichromophoric master equation}\label{a.mastereq}

General expressions for the multichromophoric rates have been obtained in Refs. \cite{Jang:2004tt,Sumi:1999ws}.
However, no general formal derivation of how MCFT can result from a full quantum approach is available in the literature. Indeed, the proof performed in Ref. \cite{Jang:2004tt} is strictly valid only for a system composed of two aggregates, a donor and an acceptor, and does not take into account the back-transfer from the acceptor to the donor.
In this section we complete our theoretical description of MCFT by deriving the master equation of a multichromophoric system using a time-convolutionless projection operator formalism:
\cite{Shibata:1977cw,Breuer:2002wp}
\begin{align}
	\ddt \mathcal P \rho_{eb}(t)
	=
	\mathcal K(t) \mathcal P \rho_{eb}(t) + \mathcal I(t)\mathcal Q\rho_{eb}(0),
	\label{e.projope}
\end{align}
where $\mathcal Q\equiv 1-\mathcal P$: the dynamics of the multichromophoric populations can be obtained from the complete density matrix $\rho_{eb}(t)$ by tracing out the irrelevant degrees of freedom by means of the projection operator $\mathcal P$.
We use the notation of Section \ref{s.rates} and set
\begin{gather}
  \mathcal P \rho_{eq}(t) = \sum_{N,n}P_n^N(t)
  \,\Pi^N_n\otimes \rho_b~, \quad \Pi^N_n = \ket{N,n}\bra{N,n}~,
  \nonumber\\
  P^N_n(t) = \Tr_b\left[\bra{N,n}\rho_{eb}(t) \ket{N,n}\right]~.
  \label{e.P}
\end{gather}
Using a second order approximation with respect to the electronic interaction between different aggregate one can show that \eqref{e.projope} reduces to
\begin{align}
  \ddt P^N_n(t) =& \sum_{M,m} k^{NM}_{nm}(t)\,P^M_m(t) - k^{MN}_{mn}(t) \, P^N_n(t)
  \nonumber
  \\&+
  \mathcal I(t)\mathcal Q\rho_{eb}(0)~,
  \label{e.pauli0}
\end{align}
with the time-dependent rates
\begin{align}
  k_{nm}^{NM}(t) =& \frac2\hbar \Re \int_0^t ds\,W^{NM}_{nm}(t,t-s)~,
  \label{e.kratenm}
  \\W^{NM}_{nm}(t,s) =&\Tr\left(\mathcal U_{-t}[J]\,\Pi^N_n\,\mathcal U_{-s}[J]\,\Pi^M_m
  \otimes \rho_b\right), \nonumber
\end{align}
where
$\mathcal U_t[\cdot] = e^{-\ih t \mathcal L}(\cdot)$, $\mathcal L(\cdot) = \sum_N [H^N_{eb}, \cdot]$ and $J=\sum_{NM} H^{NM}$.
In the following we show the further approximations needed to bring Eq.~\eqref{e.pauli0} to the form of a Pauli master equation for aggregates, i.e.
\begin{align}
	\ddt P^N(t) =& \sum_{M} k^{NM} P^M(t) - k^{MN} P^N(t),
	\label{e.pauli}
\end{align}
with $P^N(t) = \sum_n P^N_n(t)$.
The inhomogeneous term in Eq.~\eqref{e.pauli0} can be removed formally only if $\mathcal Q \rho_{eq}(0) = 0$, namely only if the excitation is localized into a single chromophore at $t=0$.
If on the other hand the excitation is initially distributed amongst different excitons, even within a single aggregate, then an inhomogeneous term is needed.
In this paper however we do not discuss in detail such an occurrence, as it is expected to affect only the dynamics of the first femtoseconds.

The stationary rates are obtained by taking the limit $t\rightarrow \infty$, assuming ergodicity. \cite{Jang:2004tt}
The multichromophoric rates then arise by summing Eq.~\eqref{e.pauli0} for different $n$ and assuming that the resulting rates are independent on $m$.
Indeed,
\begin{align}
  W^{NM}_m(s) &= \lim_{t\rightarrow\infty} \sum_n W^{NM}_{nm}(t,t-s)
  \nonumber\\ &=\Tr\left(H^{MN}\,\mathcal U_{s}[H^{NM}]\,\rho^M_{eb}[m]\right)
  \label{e.Wratem}
\end{align}
with the equilibrated state
\begin{align}
  \rho^M_{eb}[m] &\equiv
  \mathcal U_{t\rightarrow\infty}\left[\Pi^M_m \otimes \rho_b\right].
\end{align}
If the above state is independent on $m$, i.e. if the state of the aggregate $M$ equilibrates to a state which is independent on the initial position of the excitation, then the multichromophoric rate in Eq.~\eqref{e.mcrate_t} follows by taking the limit $t\rightarrow\infty$ in Eq.~\eqref{e.kratenm} owing to Eq.~\eqref{e.Wratem}, and assuming that the bath acts independently on the different chromophores.

\section{Time-convolution formalism}\label{a.tc2}

\begin{figure}[t!]
  \begin{center}
    \includegraphics[width=.3\textwidth]{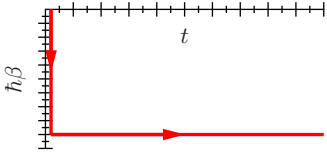}
  \end{center}
  \caption{Complex contour $\mathcal C$
  for the evaluation of the emission lineshape}
  \label{fig.contour}
\end{figure}

The time-convolution master equation is one of the standard methods for describing non-Markovian open quantum systems.
It has been derived using projection operator techniques by Nakajima \cite{Nakajima:1958to} and Zwanzig \cite{Zwanzig:1960vw} and has been applied \cite{Jang:2004tt} for the evaluation of the LSOs in Eqs.~\eqref{e.A} and \eqref{e.F}.
Owing to the cumulant expansion formalism, \cite{Kubo:1963hj,Mukamel:1979ti} it can be shown that
a second order time-convolution master equation (TC2) for the absorption lineshape operator can be derived from \eqref{e.formaleqmot} by choosing the so-called chronological ordering prescription:
\begin{align}
  	\ddt A^{\rm TC}(t)
	=
	-\ih H \, A^{\rm TC}(t) +
  	\int_0^t ds\,  K^{\rm TC}_{(2)}(t-s)\,  A^{\rm TC}(s),
  	\label{e.Anz_t}
\end{align}
where $ K^{\rm TC}_{(2)}(t-s) = e^{-\ih Ht} K(t,s) e^{\ih H s}$, i.e.
\begin{align}
  	K^{\rm TC}_{(2)}(t) = -\frac1{\hbar^2}\, C(t)\sum_a \ket a \bra a \,
  	e^{-\ih Ht}\,\ket a \bra a~.
  	\label{e.kerntc}
\end{align}
as $V_a \equiv \ket a \bra a$ where $\ket a$ represent the excited state localized in the chromophore of site $a$ in the acceptor aggregate $A$.
Similarly,
\begin{align*}
  	\ddt A^{\rm TC}(t-i\hbar\beta) =& -\ih H \, A^{\rm TC}(t-i\hbar\beta) \\&+
  	\int_0^{t-i\hbar\beta} dz\,  K^{\rm TC}_{(2)}(t-i\hbar\beta-z)\,
  A^{\rm TC}(z)
\end{align*}
so that, owing to Eq. \eqref{e.idFA}, the emission LSO can be evaluated by decomposing the above integral along the contour of Fig.~\ref{fig.contour} as
\begin{align}
  	\ddt F^{\rm TC}(t)
  	= -\ih H\, F^{\rm TC}(t) + \int_0^t ds\,
   	K^{\rm TC}_{(2)}(t-s)  \,F^{\rm TC}(s)+ I_{(2)}(t),
  	\label{e.Fnz_t}
\end{align}
where $I_{(2)}(t) = \frac{i}{\hbar Z} \int_0^\beta d\tau \, K^{\rm TC}_{(2)}\big(t-i\hbar(\beta-\tau)\big) \, e^{-\tau H}$.
On the other hand, the initial state $F(0) = A(-i\hbar\beta)/Z$ is evaluated using a second order expansion of Eq. \eqref{e.A}.
Equations \eqref{e.Anz_t} and \eqref{e.Fnz_t} were introduced in Ref.~\cite{jang:9312} for obtaining the multichromophoric LSO.
They are solved with the Fourier-Laplace transform as
\begin{align}
  	\hat A^{\rm TC}_{(2)}(\omega)
  	&=
  	-2\Im\left[\frac1{\omega-H/\hbar-i
 	\tilde K^{\rm TC}_{(2)}(\omega)}\right],
  	\label{e.Anz}
  	\\
  	\hat F^{\rm TC}_{(2)}(\omega)
	&=
	-2\Im\left[\frac{ F_{(2)}(0) +
  	\tilde I_{(2)}(\omega)}{\omega-H/\hbar-i
  	\tilde K^{\rm TC}_{(2)}(\omega)}\right],
  	\label{e.Fnz}
 \end{align}
where we used the notation $\tilde f(\omega) = \int_0^\infty dt\,e^{i\omega t}f(t)$, and $F_{(2)}(0)$ is second order expansion of the equilibrated state.
The lineshapes obtained with time-convolution master equations take the form of a Lorentzian with frequency-dependent damping kernel $\tilde K^{\rm TC}_{(2)}(\omega)$.
The above LSOs are not diagonal in the exciton basis in general because of $\tilde K^{\rm TC}_{(2)}(\omega)$ and $\tilde I_{(2)}(\omega)$: in particular the dissipation kernel $\tilde K^{\rm TC}_{(2)}(\omega)$ is diagonal in the site basis.
(See Eq.~\eqref{e.kerntc}.)
As the strength of both the dissipation kernel and of the inhomogeneous term depends on $\lambda$, and increases for increasing $\lambda$, the off-diagonal part of the lineshape is expected to give a relevant contribution in the intermediate and strong coupling regime.

Note that the emission lineshape can be obtained also from Eq.~\eqref{e.Anz} once the analytic continuation identity, Eq.~\eqref{e.idFAw}, is used.
We call
\begin{align}
  \hat F^{\rm TC+AC}_{(2)}(\omega) & \propto e^{-\beta\hbar\omega}\,
  \hat A^{\rm TC}_{(2)}(\omega)
  \label{e.Fnzsumi}
 \end{align}
such a TC2 expression.
The latter is more easy to evaluate as the second order equilibrated state and the inhomogeneous term dot not enter in the expression.
As shown in Sec.~\ref{s.tctcl}, Eqs.~\eqref{e.Fnz} and \eqref{e.idFAw} with Eq.~\eqref{e.Anz} give the same results when the second order approximation is justified.

In Eq.~\eqref{e.monomer}, the unfactorized initial condition is the origin of the stokes shift.
On the other hand, in the time-convolution master equation it introduces the inhomogeneous term.
The role of $\tilde I_{(2)}(\omega)$ in Eq.~\eqref{e.Fnz} can be better understood by using the identity $1+x\approx (1-x)^{-1} + \mathcal O(x^2)$,
\begin{align}
  	\hat F^{\rm TC}_{(2)}(\omega)
	&\simeq
	-2\Im\left[\frac{F(0)}{\omega-\left(H - \tilde S(\omega)\right)/\hbar-i \tilde K^{\rm TC}_{(2)}(\omega)}\right],
  	\label{e.Fnz_stokes}
\end{align}
where $\tilde S(\omega) = (\hbar\omega-H) {\tilde I_{(2)}(\omega)}/{F(0)}$.
The above expression elucidate the role of the inhomogeneous term; indeed one can see that the real part of $\tilde I_{(2)}$ contributes to the shift of the the eigenfrequencies of the system (Stokes shifts), while the imaginary part contributes to both the width and the envelope of the lineshape.

It is important to stress that Eqs.~\eqref{e.Anz} and \eqref{e.Fnz} do not reduce to Eq.~\eqref{e.monomer} when the aggregate consists of a single monomer.
The following section introduces a master equation for LSOs where such a problem does not subsist.

\section{Time-convolutionless formalism}\label{a.tcl2}

Time-convolutionless master equations can be obtained from the generalized cumulant expansion
by taking the so called partial ordering prescription. \cite{Kubo:1963hj,Mukamel:1979ti,VanKampen:1974wo}
In another widely used formalism they are obtained from time-convolution master
equations formally resuming the perturbative series. \cite{Shibata:1977cw}
Using a second order
time-convolutionless approach for the absorption lineshape operator, one can show that
Eq.~\eqref{e.formaleqmot} reduces to
\begin{align}
  	\ddt A^{\rm TCL}(t) &= \left(-\ih H  + \int_0^t ds\,
   	K^{\rm TCL}_{(2)}(s) \right) A^{\rm TCL}(t),
  	\label{e.Atcl}
\end{align}
where $K^{\rm TCL}_{(2)}(t-s) = e^{-i Ht/\hbar} K(t, s) e^{i Ht/\hbar}$, i.e.
\begin{align}
  	K^{\rm TCL}_{(2)}(t) = -\frac1{\hbar^2}\, C(t)\sum_a \ket a \bra a \,
  	e^{-\ih Ht}\,\ket a \bra a\,e^{\ih Ht}.
  	\label{e.kerntcl}
\end{align}
However, in this framework, the implementation the second order perturbation approach
for the emission lineshape operator results in the appearance of an inhomogeneous term
that describe the existence of unfactorized initial states, Eq.~\eqref{e.F}
\begin{align*}
  	\ddt F^{\rm TCL}(t)
	&=
	-\ih H\, F^{\rm TCL}(t)
	\\&
	\phantom{= -}+ \int_0^t ds\,K^{\rm TCL}_{(2)}(t-s)  \,F^{\rm TCL}(t)+ I^{\rm TCL}_{(2)}(t).
\end{align*}
From this expressions it is clear that, while the absorption LSO reduces to Eq.~\eqref{e.monomer} in the monomer limit, the same does not take place for the emission lineshape.
Indeed, the above equation can not reproduce the monomer case
 \begin{align}
    	\ddt F_{\rm mon}(t) &= -\left[\ih (\epsilon-2\lambda)-\dot g(t)^*\right]\,
    	F_{\rm mon}(t)
  	\label{e.monomerdiff}
\end{align}
as, in general, $I^{\rm TCL}_{(2)}(t) \neq (i/\hbar) 2\lambda F_{\rm mon}(t)$.
In order to overcome this problem the emission lineshape is obtained from the absorption lineshape
via the analytic continuation identity \eqref{e.idFA}:
\begin{align}
  	\ddt F^{\rm TCL}(t)
	&= \left(-\ih H + \int_0^{t-i\hbar\beta} dz\,
   	K^{\rm TCL}_{(2)}(z) \right)F^{\rm TCL}(t).
  	\label{e.Ftcl}
\end{align}
In the above formulation the effect of the unfactorized equilibrated state is accounted by a different {\it homogeneous} terms.
This is the prerequisite for obtaining a structure that satisfies Eq.~\eqref{e.monomerdiff} in the single monomer case: it is indeed straightforward to show that the solution of Eqs.~\eqref{e.Atcl} and \eqref{e.Ftcl} reduces to Eq.~\eqref{e.monomer} for monomer aggregates, so that they offer a more natural expression for the MC-LSOs.
They have been derived using the analytic continuation identity Eq.~\eqref{e.idFA} and the second-order time-convolution-less (TCL2) master equation.

The TCL2 lineshape expression can be better understood after some transformations.
In the exciton basis the TCL2 dissipation kernel in Eq.~\eqref{e.kerntcl} reads
\begin{align}
  K^{\rm TCL}_{(2)}(t) &= -\frac1{\hbar^2}\,
  \sum_{k,k',k''} \Xi_{k,k'}^{k''}\, C(t) \,e^{-it\omega_{k'',k'}}
  ~\ket{\varepsilon_k}\bra{\varepsilon_{k'}}
  \label{e.kerntcl_ex}
\end{align}
where $\omega_{k'',k'} = \omega_{k''}-\omega_{k'}$, $\Xi_{k,k'}^{k''}=\sum_a U_{ak}U_{ak''}^2U_{ak'}$ and $U$ is the unitary matrix of the basis transformation.
Inserting the identity one has that
\begin{align}
  \sum_{k''} \Xi_{k,k'}^{k''} \,e^{-it\omega_{k'',k'}}
  = \delta_{kk'} +
  \sum_{k''} \Xi_{k,k'}^{k''} \,\left(e^{-it\omega_{k'',k'}}-1\right)
\end{align}
and it is simple to show that Eq.~\eqref{e.Atclfact} holds, namely that $A^{\rm TCL}(t) = a_{ex}(t) \cdot a(t)$, where $a_{ex}(t)=e^{-i Ht/\hbar-g(t)}$ and $a(t)$ is the matrix solution of the differential equation, Eq.~\eqref{e.Lope}.

\section{Correlation function}
For the Drude-Lorentz spectral density it is known that
  \begin{align}
    C(t)
    = \sum_{k=0}^{\infty} c_k\,e^{-\nu_k\,t},
  \end{align}
  where
  \begin{align}
    c_0
    &=
    \hbar\lambda\gamma\left(\cot\left(\frac{\beta\hbar\gamma}2\right)
    -i\right),
    &
    \nu_0 &= \gamma,
    \\
    c_k
    &=
    \frac{4\lambda}{\beta} \frac{\gamma\nu_k}{\nu_k^2-\gamma^2},
    & \nu_k &= \frac{2\pi k}{\beta\hbar}.
    \label{e.dlparam}
  \end{align}
The other quantities such as the line-broadening function $g(t)$ or the operator elements $O(t)$ in Eq.~\eqref{e.Lope} are then obtained from the above expressions. The results are formally summed with a computer algebra system.
It is found that they can be written as a combination of hypergeometric and polygamma functions.
The latter are then implemented for the numerical simulations.


\end{document}